\documentclass[aps,prl,floatfix,twocolumn,groupedaddress,showpacs]{revtex4}

\usepackage{graphicx}
\usepackage{amsmath}
\usepackage{amssymb}
\usepackage{amsthm}
\usepackage{array}
\usepackage{xy}
\usepackage{subfigure}

\newcommand{\gtsimeq}{\raisebox{-0.6ex}{$\,\stackrel
        {\raisebox{-.2ex}{$\textstyle >$}}{\sim}\,$}}

\begin{document}


\title{Activated dynamics and effective temperature in a steady state sheared glass}


\author{Thomas K. Haxton}
\author{Andrea J. Liu}
\affiliation{Department of Physics and Astronomy, University of Pennsylvania, Philadelphia, PA 19104}


\date{\today}

\begin{abstract}

We conduct nonequilibrium molecular dynamics simulations to measure the shear stress $\sigma$, the average inherent structure energy $\langle E_{\rm IS}\rangle$, and the effective temperature $T_{\rm eff}$ of  a sheared model glass as a function of bath temperature $T$ and shear strain rate $\dot \gamma$.    For $T$ above the glass transition temperature $T_0$, the rheology approaches a Newtonian limit and $T_{\rm eff} \rightarrow T$ as $\dot{\gamma} \rightarrow 0$, while for $T<T_0$, $\sigma$ approaches a yield stress and $T_{\rm eff}$ approaches a limiting value near $T_0$.  In the shear-dominated regime at high $T$, high $\dot \gamma$ or at low $T$, we find that $\sigma$ and $\langle E_{\rm IS}\rangle$ each collapse onto a single curve as a function of $T_{\rm eff}$.  This indicates that $T_{\rm eff}$ is controlling behavior in this regime.  

\end{abstract}


\maketitle


When a liquid is quenched through the glass transition temperature $T_0$, it falls out of equilibrium and becomes a glass.  Below $T_0$, the thermal energy is insufficient for the system to surmount energy barriers on accessible time scales, so the glass explores relatively few  configurations.  However, if the glass is held in contact with a thermal reservoir and sheared at a fixed rate, it can reach a steady state in which it explores many different minima in the energy landscape, even if the temperature of the reservoir is well below $T_0$~\cite{Kurchan2001}.  The steadily sheared glass is far from equilibrium: energy is continually supplied on long time and length scales via the boundaries, and is removed on short scales by the thermal reservoir.   Nonetheless, simulations show that fluctuations in such systems are well described by an effective temperature $T_{\rm eff}$ that is higher than the bath temperature~\cite{Barrat2000,OHern2004,Ilg2007}, as predicted theoretically~\cite{Cugliandolo1997}.   Nine different definitions yield a common value of $T_{\rm eff}$~\cite{Makse2002,Barrat2000,Ono2002,OHern2004,Ilg2007,Danino2005}, providing strong numerical evidence for the utility of the concept.  Most recently, Ilg and Barrat~\cite{Ilg2007} showed that $T_{\rm eff}$ controls activated transition rates of a test probe consisting of a dimer connected by a double-well potential, embedded in a sheared model glass.  The rate of crossing the energy barrier separating the two wells has an Arrhenius form $\exp (-\Delta E/T_{\rm eff})$, where $T_{\rm eff}$ is consistent with previous definitions.

In this paper, we shift the focus from testing the validity of effective temperature to examining its importance for materials properties.  We show that $T_{\rm eff}$ plays a critical role in fluidizing a glass.  We find that the shear stress collapses onto a single curve depending only on $T_{\rm eff}$ whenever the shear rate is high enough to dominate over thermal effects.  The average inherent structure energy collapses in similar fashion.  These findings suggest that $T_{\rm eff}$ activates particle rearrangements necessary for flow, much as thermal fluctuations do in an equilibrium liquid, supporting the idea that a common framework might describe unjamming by mechanical forcing and by temperature~\cite{Liu1998}.

Our simulation model is a two-dimensional glass-forming liquid composed of disks interacting via a purely repulsive harmonic potential~\cite{OHern2004}.  We study 50:50 mixtures of disks of diameter ratio $1:1.4$ and equal mass.  The area fraction is fixed at $\phi=0.9$.  Most of the results are based on simulations of $400$ disks, but we carried out simulations of up to 6400 disks to confirm that none of these results have any appreciable system size dependence.  Units in this paper are measured with the smaller particle diameter, the interaction spring constant, the particle mass, and the Boltzmann factor set equal to 1.  This yields a unit time period on the order of a binary collision time.  We uniformly and steadily shear the system at a strain rate $\dot \gamma$ and couple the system to a heat bath at temperature $T$ by solving the Sllod equations of motion with Lees-Edwards periodic boundary conditions and a Gaussian thermostat~\cite{Allen1987}.  We integrate these equations using a fourth-order Gear corrector-predictor algorithm with a time step of $0.01.$  We obtain the same results with a Nos\'e-Hoover thermostat.  We use between five and twenty simulation runs for each set of $(T,\dot{\gamma})$.  For each simulation, we collect data over at least nine strain units after an equilibration period of several strain units. 

We measure $T_{\rm eff}$ from the relation between the static linear response and the variance of the pressure~\cite{Allen1987, Ono2002}.  In equilibrium at fixed $N, T$, and $A$, this relation is 
\begin{equation}\label{tp}\dfrac{A}{T}\langle(\delta p)^{2}\rangle=A\left(\dfrac{\partial{\langle p\rangle}}{\partial{A}}\right)_T+\langle p \rangle+\dfrac{\langle x \rangle}{A},\end{equation}
where $p$ is the pressure, $A$ is the area, and $x$ is the hypervirial as defined in \cite{Allen1987}.  $T_{\rm eff}$ is defined by replacing $T$ with $T_{\rm eff}$ in the left hand side of Eq.~(\ref{tp}).  We measure $\partial \langle p\rangle/\partial A$ by running simulations at $\phi=0.897$ and $\phi=0.903$, using the same protocol and a similar quantity of simulations as for $\phi=0.9$.  Measurements from this definition have been compared to those from many other definitions of $T_{\rm eff}$ for a zero-temperature sheared foam~\cite{Ono2002}.  We find that $T_{\rm eff}$ is consistent with less precise measurements of $T_{\rm eff}$ defined by the Green-Kubo relation for shear viscosity~\cite{Ono2002, Allen1987}.   O'Hern et al.~\cite{OHern2004} showed that $T_{\rm eff}$ from pressure fluctuations agrees with that derived from the time-dependent linear response of density fluctuations~\cite{Barrat2000} over a range of parameters for the system we use.  We also find consistency between these two definitions, except at very low strain rates where the logarithmic time-dependence of the diffusivity, expected in two dimensions~\cite{Kawasaki1973}, is apparent within the time scale of the density correlation function.  This long-time tail does not affect the viscosity because at the high area fraction and low temperatures studied, the kinetic contribution to the viscosity is much smaller than the potential contribution.

\begin{figure}
\includegraphics[width=8.5cm]{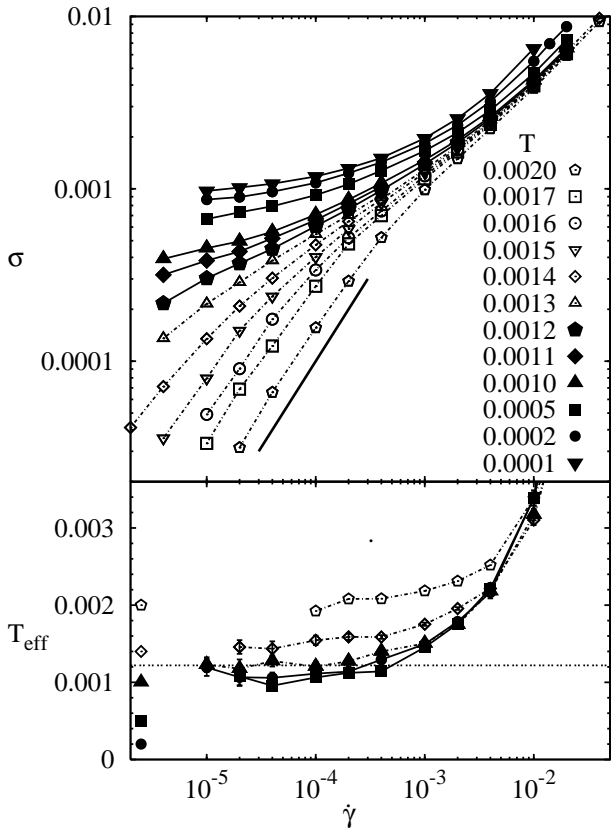}%
\caption{$\sigma$ (a) and $T_{\rm eff}$ (b) \textit{vs.} $\dot \gamma$ for several values of $T$ and $\dot\gamma$.  In all figures, dashed lines and open points indicate $T>T_0$.   Solid lines and solid points indicate $T<T_0$.  In (a), the straight line indicates a slope of 1, expected for a Newtonian fluid.  In (b), the horizontal line indicates $T_0$, while the isolated points near the left axis indicate the values of $T$.}
\label{results}
\end{figure}

Figure~\ref{results} shows the measured quantities, stress $\sigma$ and effective temperature $T_{\rm eff}$, as functions of  bath temperature $T$ and strain rate $\dot \gamma$.   Figure~\ref{results}(a) shows a bifurcation of $\sigma$ about a critical temperature $T_0=0.0012$ at low $\dot{\gamma}$, similar to what has been observed in experiments~\cite{Lu2003} and earlier simulations~\cite{Barrat2000} of sheared glass.  Bifurcations about a critical volume fraction have been observed in experiments on colloids~\cite{Senff1999} and emulsions~\cite{Mason1996} and in simulations of frictionless spherical packings~\cite{Olsson2007}.   At $T_0$, the stress obeys power-law scaling with $\dot\gamma$, with $\sigma \sim \dot\gamma^{0.3}$.

The two sides of the bifurcation in Fig.~\ref{results}(a) describe two different low-$\dot\gamma$ limits.  On the $T>T_0$ side, $\sigma$ approaches $\sigma \sim \dot \gamma$ at low $\dot\gamma$, suggesting that the shear viscosity $\eta \equiv \sigma/\dot\gamma$ enters a Newtonian regime for $T>T_0$ and sufficiently low $\dot\gamma.$  For $T\ge 0.0017$, we have reached strain rates low enough that $\eta$ becomes independent of $\dot\gamma$.  We define the equilibrium viscosity $\eta_{\rm eq}$ as the shear viscosity in this Newtonian regime.  For $T_0<T<0.0017$, we do not reach the Newtonian regime at accessible time scales/strain rates.  However, we find that for all $T\ge0.0015$, $\eta(\dot\gamma)$ is well-described by the phenomenological Ellis equation~\cite{Xu2005}, $1/\eta(\sigma)=1/\eta_0+m^{-1/n}\sigma^{(1-n)/n}$,
which interpolates between Newtonian and power law rheology and allows us to define $\eta_{\rm eq}$ down to $T=0.0015$.  

In contrast, for $T<T_0$ we observe apparent yield stress rheology on the time scale of our simulations~\cite{Sollich1997}.  We find $\sigma-\sigma_{\rm yield}\propto \dot{\gamma}^b$ over the lowest two decades of $\dot{\gamma}$ that our simulation can access, with the exponent $b$ ranging between $0.4$ $(T=0.0012)$ and $0.6$ $(T=0.0001)$, similar to that observed in emulsions~\cite{Mason1996}.  This implies that the viscosity diverges as $\eta=\sigma/(\sigma-\sigma_{\rm yield})^{1/b}$ as $\sigma \rightarrow \sigma_{\rm yield}$.  

Figure~\ref{results}(b) shows the dependence of $T_{\rm eff}$ on  $T$ and $\dot \gamma$.  For all $T$, $T_{\rm eff}$ approaches a limiting value in the quasistatic limit, $\dot \gamma \rightarrow 0$.  For $T>T_0,$ that limiting value is simply $T$.  However, for $T<T_0$, $T_{\rm eff}$ appears to saturate to a value $T_{\rm eff, 0}$ near $T_0$~\cite{Ono2002}.  Such a saturation of $T_{\rm eff}$ at low $\dot \gamma$ has been observed in experiments on sheared granular packings~\cite{Song2005, Corwin2005}.  This apparent quasistatic limit suggests that $T_{\rm eff, 0}$ is a property of the unsheared glass at bath temperature $T$, describing the disorder associated with different minima in the energy landscape~\cite{Kob2000}.  

\begin{figure}
\includegraphics[width=8.5cm]{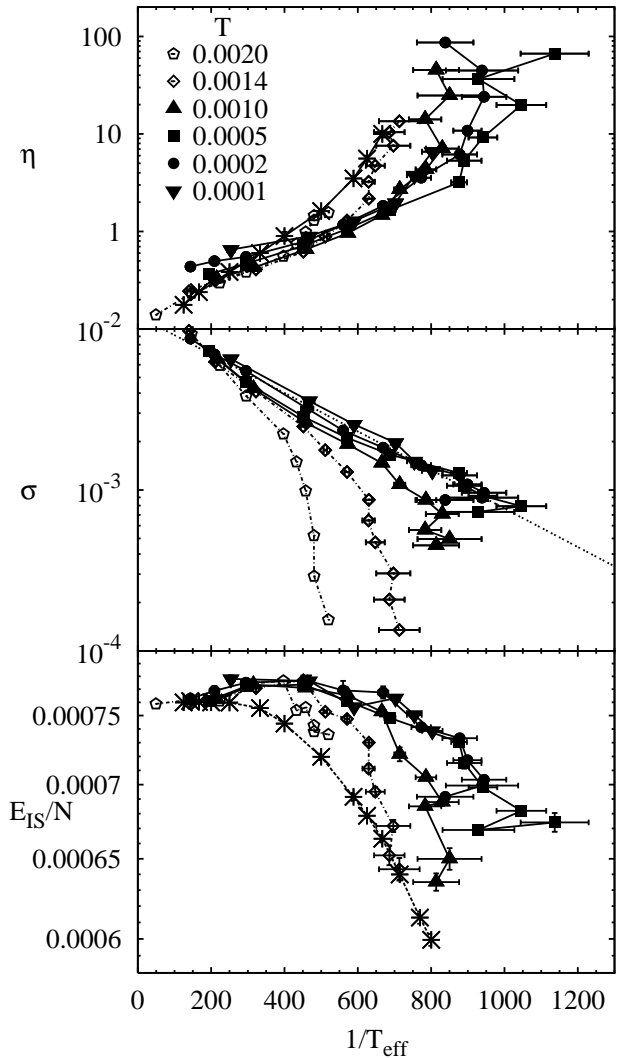}%
\caption{(a) $\eta$ \textit{vs.} $1/T_{\rm eff}$ for several values of $T$ and $\dot\gamma$.  Asterisks correspond to the Newtonian viscosities ($\dot\gamma\rightarrow 0$, $T_{\rm eff}=T$).  (b) $\sigma$ \textit{vs.} $1/T_{\rm eff}$.  The straight dotted line is a fit to the form $\sigma=\sigma_0\exp(-\Delta E/T_{\rm eff})$ for the three lowest bath temperatures.  (c) $\langle E_{\rm IS} \rangle$ \textit{vs.} $1/T_{\rm eff}$.  Asterisks indicate $\langle E_{\rm IS} \rangle$ for $\dot\gamma=0$ ($T_{\rm eff}=T$).  \label{multi}}
\end{figure}

In Fig.~\ref{multi}(a), we compare the approach to jamming as $T\rightarrow T_0$ with the approach at fixed $T<T_0$ and $\dot \gamma \rightarrow 0$, parameterizing the latter approach by $T_{\rm eff}$ rather than $\dot \gamma$.  The dependence of $\eta$ on $1/T_{\rm eff}$ is  
similar to that of $\eta_{\rm eq}$ on $1/T$~\cite{Langer2000},
but we find no collapse among the different approaches to jamming.  Along the equilibrium approach, $\eta$ has the Arrhenius form $\eta=\eta_\infty \exp(E_A/T)$ at high $T$.  The non-equilibrium approaches exhibit no Arrhenius regime in $\eta$.  However, another reasonable measure of relaxation time, $\tau_{\rm shear}\equiv\dot\gamma^{-1}$ (not shown), does vary in Arrhenius fashion with $T_{\rm eff}$ at high $T_{\rm eff}$.  
Along all approaches, $\eta$ is super-Arrhenius for $T_{\rm eff}$ near $T_0$ or $T_{\rm eff, 0}$.

Figure~\ref{multi}(b) demonstrates that the stress $\sigma$ collapses as a function of $T_{\rm eff}$ in the shear-dominated regime.   We find that $\sigma$ varies between two limits, depending on $T$ and $\dot \gamma$.  For $T>T_0$ and $\dot{\gamma} \rightarrow 0$, the shear stress approaches the Newtonian limit, $\sigma= \eta_{\rm eq}\dot{\gamma}$, while $T_{\rm eff}$ approaches $T$.   In this regime, thermal fluctuations $T$ dominate over shear-induced ones.  However, for $T\ll T_0$ and any $\dot \gamma$, or for $T\gtsimeq T_0$ and high $\dot{\gamma}$, the system crosses over to another regime, where shear-induced fluctuations dominate over thermal ones.  In this shear-driven regime, $\sigma$ depends on $T$ and $\dot{\gamma}$ only through $T_{\rm eff}$.  The dependence of $\sigma$ on $T_{\rm eff}$ follows the simple form
\begin{equation}\label{eqn}
\sigma\approx\sigma_0\textrm{exp}(-\Delta E/T_{\rm eff}),
\end{equation}
represented by a dotted line in Fig.~\ref{multi}(b).   The collapse of the data from different $T$ and $\dot \gamma$ onto this same curve indicates that the dynamics are most properly described as being controlled by $T_{\rm eff}$, not $\dot \gamma$ or $T$.  In the shear-dominated regime, the height of the energy 
scale, $\Delta E=0.0028\pm 0.0001$, is independent of $T$ and $\dot \gamma$.  Its value is comparable to the activation energy $E_A=0.0023 \pm 0.0002$ extracted from the high-$T$ equilibrium Arrhenius relationship $\eta=\eta_\infty \exp(E_A/T)$.  
The correspondence between $\sigma$ and $T_{\rm eff}$ suggests that the existence of a finite yield stress would imply a nonzero quasistatic value of $T_{\rm eff, 0}$.  

Equation~(\ref{eqn}) suggests a
simple scenario.  Suppose that the power per unit area supplied globally by shearing, $\sigma\dot{\gamma}$, were used to overcome local energy barriers of height $\Delta E$ at a rate of $R$ transitions per unit area per unit time.  Using $\sigma \dot \gamma=R\Delta E$ and the observed Eq.~(\ref{eqn}), we find
\begin{equation}
 R=R_0\dot\gamma \textrm{exp}(-\Delta E/T_{\rm eff}), \label{rdef}
 \end{equation}
with $R_0=\sigma_0/\Delta E=4 \pm 1$.  This implies that the rate of barrier crossing is Arrhenius in $T_{\rm eff}$ with an attempt frequency proportional to the strain rate $\dot \gamma$.  Local rearrangements whose rate scales with $\dot \gamma$ are observed in foam experiments~\cite{Gopal1999} and simulations~\cite{Tewari1999}, as well as in athermal quasistatic simulations of soft disks~\cite{Maloney2004}.

Although it was derived from a steady-state relationship, Eq.~\ref{rdef} motivates a hypothesis for the time evolution of $T_{\rm eff}$. Such equations~\cite{Lemaitre2002,Langer2004} have been used in the context of shear transformation zone (STZ) theory~\cite{Falk1998}.
Following previous work~\cite{Lemaitre2002}, we suppose that the equation should balance a heating rate proportional to the work done on the system with a relaxation rate proportional to $\exp(-E_1/T_{\rm eff})$.  As in Eq.~\ref{rdef} and differing from \cite{Lemaitre2002}, we suggest that the relaxation rate should depend on a scalar rate at which the system explores new configurations.  These considerations yield
\begin{equation}\label{tom}
\dot T_{\rm eff}\propto Q-\nu \exp(-E_1/T_{\rm eff}),
\end{equation} 
where $Q \propto \sigma \dot \gamma$~\cite{Langer2003} and the attempt frequency $\nu \propto \dot \gamma$ in the case of steady-state shear.  Equation~\ref{tom} is consistent with Eq.~\ref{eqn} in the steady-state limit.  For $T_{\rm eff}$ near its steady-state value $T_{\rm eff,SS}$, Eq.~\ref{tom} reduces to 
$\dot T_{\rm eff}\propto \sigma\dot\gamma (T_{\rm eff, SS}-T_{\rm eff})$, as used in STZ theory~\cite{Langer2004}.  

The scenario suggested by Eqs.~\ref{rdef}-\ref{tom} is that $T_{\rm eff}$ activates the system over barriers whose height is independent of $T_{\rm eff}$.  However, measurements of the average inherent structure energy $\langle E_{\rm IS}\rangle$~\cite{Stillinger1982} of the system suggest that the height of energy barriers does depend on $T_{\rm eff}$.  We measure $E_{\rm IS}$ of the sheared system by taking configurations explored during steady-state shear and quenching them to their local energy minima by the conjugate-gradient technique.  For comparison, we also measure $\langle E_{\rm IS}\rangle(T_{\rm eff}=T)$ for the equilibrium system at $\dot\gamma=0$ above the glass transition temperature.  Fig.~\ref{multi}(c) shows that $\langle E_{\rm IS} \rangle(T_{\rm eff}=T)$ is flat at high $T$. 
For $T<T_{MC}$, where $T_{MC}$ marks the onset of super-Arrhenius behavior in Fig.~\ref{multi}(a), we find that $\langle E_{\rm IS} \rangle(T_{\rm eff}=T)$ decreases monotonically with decreasing $T$, in agreement with~\cite{Jonsson1988}.  At low $T$, the system visits deeper potential energy basins, presumably separated by higher barriers, consistent with the super-Arrhenius viscosity $\eta=\eta_{\infty}\exp(E_A(T)/T)$, where $E_A(T)$ is the $T$-dependent barrier height.

Figure~\ref{multi}(c) shows that $\langle E_{\rm IS} \rangle(T_{\rm eff})$ of the sheared system is distinct from the equilibrium curve~\cite{Kob2000} but also decreases as $T_{\rm eff}\rightarrow T_{\rm eff, 0}$.  Moreover, the data appear to collapse in the shear-dominated regime, as in Fig.~\ref{multi}(b).  At sufficiently low $T$ and/or high $\dot\gamma$, the data collapse onto a single curve, but cross over to the equilibrium curve at high $T$ and low $\dot \gamma$.  For each $T$, the point at which $\langle E_{\rm IS} \rangle(T_{\rm eff})$ begins to decrease corresponds to the upturn of viscosity, suggesting that energy barriers increase with decreasing $\langle E_{\rm IS} \rangle$.

In summary, there are two possible explanations for the upturn of $\eta$ as $T_{\rm eff} \rightarrow T_{\rm eff, 0}$.  The first explanation is that relaxation rates are controlled by $T_{\rm eff}$-activated transitions over barriers whose heights increase with decreasing $T_{\rm eff}$.
This view is supported by an STZ analysis of our data~\cite{Langer2007}.  The second explanation is that the barrier heights overcome during the shearing process do not depend strongly on $T_{\rm eff}$, even though the energy minima decrease with decreasing $T_{\rm eff}$.  In that case, the super-Arrhenius behavior is due to Eq.~\ref{eqn} and the divergence in $\eta \approx \sigma_{\rm yield}/\dot\gamma$ as $\dot\gamma \rightarrow 0$.  Further study, particularly of transients, is
needed to resolve this issue.

Finally, we revisit the issue of the validity of the effective temperature concept.  Nine independent definitions of temperature have been shown to yield consistent values of $T_{\rm eff}$, within numerical error:  the relation of density~\cite{Barrat2000,OHern2004} and pressure fluctuations~\cite{OHern2004} at nonzero wavevectors to the associated response at long times; the relation of fluctuations in the total pressure, stress, and energy to static response~\cite{Ono2002,OHern2004};  the Einstein relation between diffusion and drag~\cite{Barrat2000,Ono2002,Makse2002}; the derivative of entropy with respect to energy~\cite{Makse2002,Ono2002}; the fluctuations of a low-frequency harmonic oscillator~\cite{Danino2005}; and the barrier crossing rate of a test two-level system~\cite{Ilg2007}.   However, there are definitions that do not yield consistent values of $T_{\rm eff}$: the relation of fluctuations in the total deviatoric pressure and in the vorticity component of pressure to the response at long times~\cite{OHern2004}.  Thus, the concept of effective temperature, even when restricted to long-time-scale properties, is only approximate~\cite{Cugliandolo1997,Shokef2006}. 

If the concept of effective temperature is not rigorously valid for sheared glasses, why would it be of any interest for these systems?  Our results provide an answer:   the effective temperature critically affects materials properties by setting the energy scale for fluctuations that kick flowing glasses over energy barriers. 

We thank D. J. Durian, M. L. Falk, A. Gopinathan, K. C. Lee, M. L. Manning, S. R. Nagel, C. S. O'Hern, Y. Shokef and particularly J. S. Langer for instructive discussions, and P. Ilg and J.-L. Barrat for showing us their results before publication.  We gratefully acknowledge the support of NSF-DMR-0605044 and the hospitality of the Aspen Center for Physics.

\bibliography{arr}

\end{document}